\documentclass[prd,twocolumn,floatfix,amsmath,amssymb,floatfix,nofootinbib]{revtex4}
\usepackage{graphicx,color,dcolumn,booktabs,bm}
\usepackage{amsmath}
\usepackage{longtable,lscape,comment,braket}
\usepackage{txfonts}
\usepackage{overpic}
\usepackage{indentfirst}
\usepackage{feynmf}   
\usepackage{slashed}  
\usepackage{cases}
\usepackage{epstopdf}
\usepackage{psfrag}
\usepackage{subfigure}
\usepackage{threeparttable}
\pdfoutput=1
\usepackage{color}
\usepackage{multirow}

\graphicspath{{Figures/}}
\usepackage[colorlinks, citecolor=blue,anchorcolor=red,menucolor=red,
linkcolor=blue,filecolor=red,runcolor=red,urlcolor=blue,frenchlinks=red]{hyperref}

\usepackage{ulem}
\def\be{\begin{equation}}
\def\ee{\end{equation}}
\def\bal#1\eal{\begin{align}#1\end{align}}
\def\PPP/{${}^3P_0$}
\def\cce/{coupled-channel effects}
\def\doned/{$D_1\bar{D}$}
\def\donedstar/{$D_1\bar{D}^*$}
\def\dtwodstar/{$D_2^*\bar{D}^*$} 
\def\dtwod/{$D_2^*\bar{D}$}
\def\dmeson/{charmed meson}
\def\swave/{$S$ wave}
\def\dwave/{$D$ wave}
\def\xy/{$Y(4260)$}
\def\xyy/{$Y(4360)$}
\def\ppsi/{$\psi(4415)$}
\def\bal#1\eal{\begin{align}#1{}\end{align}}
\def\cce/{coupled-channel effects}
\newcommand{\vect}[1]{\boldsymbol{\mathbf{#1}}}
\newcommand \inner [2] {\langle {#1}\vert {#2}\rangle}

\def\be{\begin{equation}}
\def\ee{\end{equation}}
\def\ba{\begin{eqnarray}}
\def\ea{\end{eqnarray}}
\def\3P0{{}^3P_0}

\begin{document}

\title{Heavy quark spin partners of the $Y(4260)$ in coupled-channel formalism}

\author{Muhammad Naeem Anwar$^{1,2}$}\email[]{m.anwar@fz-juelich.de}
\author{Yu Lu$^{3}$}\email[]{ylu@ucas.ac.cn}

\affiliation{
$^1$Institute for Advanced Simulation, Institut f\"ur Kernphysik and J\"ulich Center for Hadron Physics, Forschungszentrum J\"ulich, D-52425 J\"ulich, Germany\\
$^2$Helmholtz-Institut f\"ur Strahlen- und Kernphysik and Bethe Center for Theoretical Physics, Universit\"at Bonn,  D-53115 Bonn, Germany\\
$^3$School of Physical Sciences, University of Chinese Academy of Sciences, Beijing 100049, China\\
}

\date{\today}

\begin{abstract}

The charmoniumlike state \xy/ is described as predominantly a \doned/ molecule in a coupled-channel quark model
[Phys.\,Rev.\,D\,\textbf{96},\,114022\,(2017)].
The heavy quark spin symmetry (HQSS) thus implies the possible
emergence of its heavy quark spin partners with molecular configurations as \donedstar/ and \dtwodstar/
below these charmed mesons' thresholds.
We analyze the probabilities of various intermediate charmed meson loops for
$J^{PC}=1^{--}$ exotic state \xyy/ and find that the channel \donedstar/ couples more strongly
around its mass regime, and the coupling behavior remains the same even if the
mass of \xyy/ is pushed closer to \donedstar/  threshold. This enlightens that the most favorable molecular
scenario for the \xyy/ could be \donedstar/, and hence it can be interpreted as HQSS partner of the \xy/.
We also find the strong coupling behavior of \dtwodstar/ channel with the $\psi(4415)$,
which makes it a good candidate for a dominant \dtwodstar/ molecule.
We discuss the important decay patterns of these resonances to disentangle their long- and short-distance structures.

\end{abstract}

\maketitle

\section{Introduction} \label{introduction}

Heavy quark spin symmetry (HQSS) implies that the hadronic interactions do not depend on the spin of the heavy quark, and heavy mesons can be fully classified by using the quantum numbers of the light quark cloud. If a hadron with a given heavy quark spin is observed experimentally, it is then natural to expect that there should exist its spin partners with different heavy quark spin but with the same light degree of freedom. Such heavy quark spin patterns are identified for instance for recently observed LHCb pentaquarks \cite{Du:2019pij,Zou:2021sha,Liu:2019tjn,Xiao:2019aya,Meng:2019ilv,Shimizu:2019ptd}, and for charged exotic states in the bottom sector, namely $Z_b$s~\cite{Bondar:2011ev,Cleven:2011gp,Baru:2017gwo,Mehen:2011yh,Ohkoda:2011vj,Sun:2011uh}.

\begin{figure}
  \centering
  \includegraphics[width=0.49\textwidth]{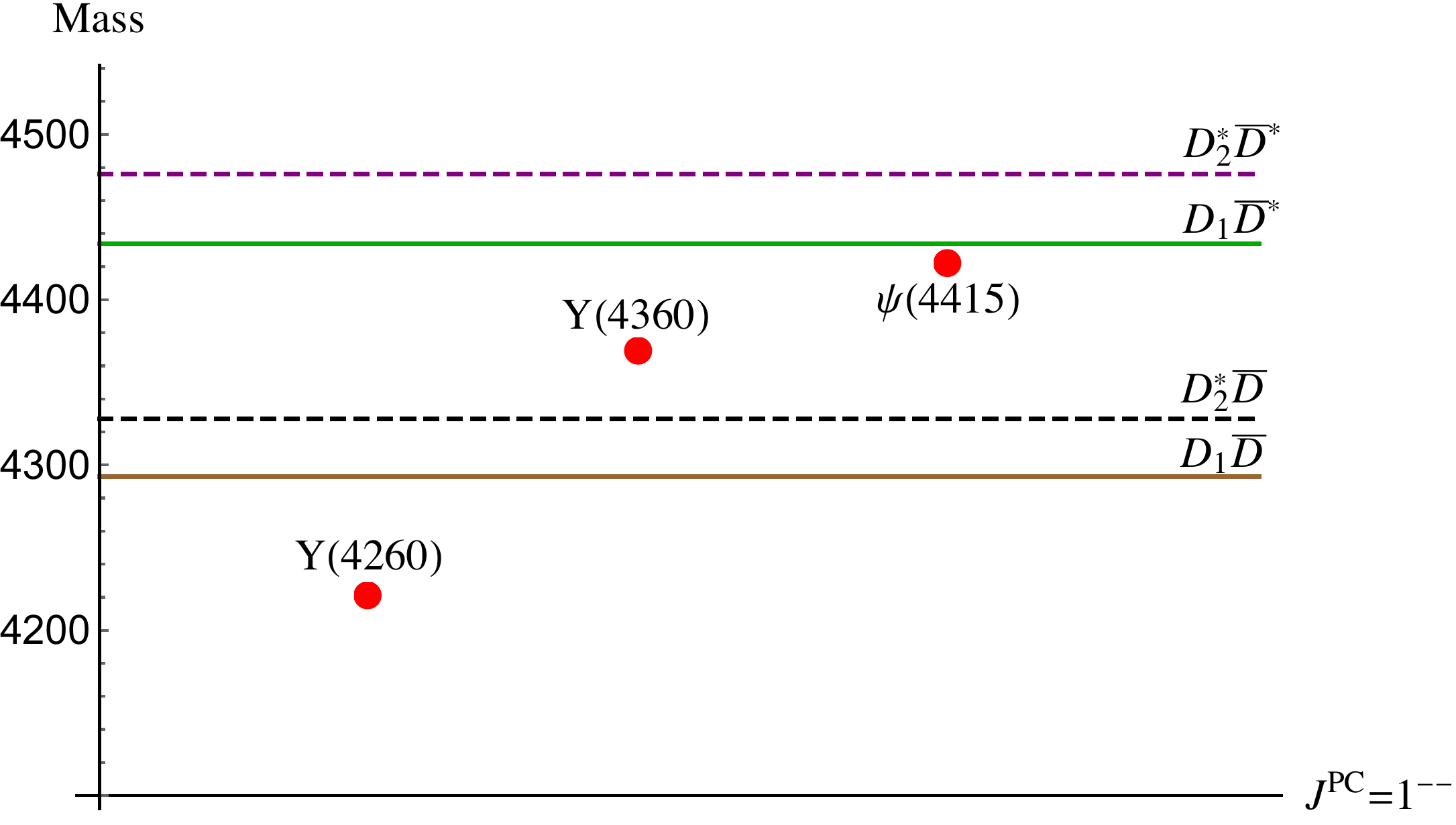}\\
  \caption{Important charmed meson thresholds in the proximity of \xy/, \xyy/ and $\psi(4415)$.}
  \label{levels.pdf}
\end{figure}

In this work, we predict the heavy quark spin partners of the exotic state \xy/, provided that this state is predominantly a \doned/ hadronic molecule\footnote{Throughout in this paper, \doned/, \donedstar/, and \dtwodstar/ mean $D_1(2420)\bar{D}$+$c.c.$,  $D_1(2420)\bar{D}^*$+$c.c.$, and  $D_2(2460)\bar{D}^*$+$c.c.$, respectively.}~\cite{Wang:2013cya}.
The HQSS thus implies the possible emergence of other hadronic molecules
such as \donedstar/ and $D^{*}_2\bar{D}^*$ below or nearby these thresholds.
We demonstrate that, as \xy/ is dominant by the \doned/ molecular component,
its HQSS partners \donedstar/ and \dtwodstar/ then naturally emerge and
we identify them as $J^{PC}=1^{--}$ charmoniumlike state \xyy/ and \ppsi/, respectively.
It is wort noting that the mass difference between \xyy/ and \xy/ is almost same
as the mass difference between vector and pseudoscalar charmed meson, namely
\be
M_{Y(4360)}-M_{Y(4260)} \simeq M_{D^{*}}-M_{D}~.
\ee
This makes a good benchmark to expect \xyy/ as an ideal candidate for the HQSS partner of the \xy/.
The nearest threshold involving two charmed mesons above \xyy/ is \donedstar/
(see Fig.\,\ref{levels.pdf} for other threshold levels), and it is approximately as far as \doned/ from the
\xy/\footnote{Note that the Particle Data Group (PDG) mass value of the \xy/ is ($4220 \pm 15$) MeV~\cite{ParticleDataGroup:2020ssz}.}.
This reflects an important consequence of the HQSS\textemdash the binding energies of these meson molecules would be of the same order~\cite{Guo:2017jvc}. Therefore, one can expect that the coupling strength of \xyy/ with the first few channels will be of the same order as we observed for the \xy/ case. We will briefly come back to this point in Sec.~\ref{results}.
Moreover, HQSS implies the presence of \doned/ and \dtwod/ components in the wave function of \xyy/,
the latter channel couples to $J^{PC}=1^{--}$ through $D$-wave only.
However, their masses are below the experimental value of the mass of \xyy/, and hence will be neglected here.
The mass values of relevant thresholds can be read from the Appendix (Table~\ref{thresholds}).

Before we go into details of our calculations, we first briefly review the status of \xyy/.
This charmoniumlike structure was first observed in 2007 at BaBar in the initial state radiation (ISR) process $e^+e^-\to  \pi^+ \pi^- \psi(2S)$~\cite{Aubert:2007zz}, and later, in the same year at Belle~\cite{Wang:2007ea}. Subsequently, BaBar and Belle have updated their data~\cite{Lees:2012pv,Wang:2014hta}. Recently, BESIII Collaboration has observed this state for the first time in the $e^+e^-\to \pi^+ \pi^- J/\psi$ process~\cite{BESIII:2016bnd}. We summarize the extractions of mass and width of the \xyy/ from different experimental data in Table~\ref{data}. More details on its status can be found in a recent review~\cite{Brambilla:2019esw}, and for different theoretical interpretations, see Sec.\,4.8 of the review article~\cite{Chen:2016qju}.

\begin{table}
\label{data}
  \renewcommand\arraystretch{1.5}
  \setlength{\tabcolsep}{8pt}
  \centering
  \begin{tabular}{cccc}
  \hline\hline
  Mass (MeV) & Width $\Gamma$ (MeV) & Measurement\\
  \hline
   $4324 \pm 24$            & $172\pm 33$                       & BaBar 2007~\cite{Aubert:2007zz}\\
  $4361 \pm 9 \pm 9$        & $74\pm 15\pm 10$                  & Belle 2007~\cite{Wang:2007ea}\\
  $4347 \pm 6 \pm 3$        & $103\pm 9\pm 15$                  & BaBar 2014~\cite{Lees:2012pv}\\
  $4340 \pm 16 \pm 9$       & $94\pm 32\pm 13$                  & Belle 2015~\cite{Wang:2014hta}\\
  $4320 \pm 10.4 \pm 7.0$   & $84.2 \pm12.5 \pm 2.1 $ & BESIII 2017~\cite{BESIII:2016bnd}\\
  $4383.8\pm 4.2 \pm 0.8$   & $101.4^{+25.3}_{-19.7} \pm 10.2 $ & BESIII 2017~\cite{BESIII:2017tqk}\\
  \hline
  $4368 \pm 13$ & $96\pm 7$ & PDG Average~\cite{ParticleDataGroup:2020ssz} \\
  \hline\hline
  \end{tabular}
\caption{Masses and widths extracted from the different experimental data for the \xyy/ in the $e^+e^-\to  \pi^+ \pi^- \psi(2S)$ process. Only the BESIII 2017 extraction~\cite{BESIII:2016bnd} is based on the $e^+e^-\to \pi^+ \pi^- J/\psi$ process.}
\end{table}

\xyy/ is recognized as a potential candidate for an exotic state since it does not show strong coupling
to open-charm channels (such as $D\bar{D}$) which are generally expected as dominant decays of vector charmonia.
Moreover, there is not any pronounced enhancement around the \xyy/ mass in the inclusive cross sections
$e^+ e^- \to \textrm{hadrons}$, the so-called $R$-value measurements.
Hence, it is necessary to make more effort to investigate the structure of \xyy/ with updated knowledge of multiquark dynamics.
We use this opportunity and report our analysis of this exotic state.

The $J^{PC}=1^{--}$ exotic state \xy/ is being analyzed in the coupled-channel quark model~\cite{Lu:2017yhl},
where the constituent quark model is used to describe the bare quark-antiquark interaction, 
and the \PPP/ quark-pair creation mechanism is used to couple charmonium core to the molecular components. 
The probabilities of the various charmed meson components of the \xy/
are analyzed and it is found that, even though the HQSS forbids $S$-wave coupling of \doned/ to the ${}^3S_1$ charmonia
[$\psi(nS)$], the $D$-wave coupling is allowed and not negligible.
The \xy/ is interpreted as a mixture of a charmonium core plus the dominant \doned/ component.
To probe the charmonium core of \xy/, the coupling behavior of the \doned/ channel with different charmonium cores
is investigated and it is found that it couples more strongly to ${}^3D_1$ charmonia [$\psi(nD)$].
This manifests that the charmonium core of the \xy/ is likely to be $\psi(nD)$.
The production of $\psi(nD)$ via $e^+ e^-$ annihilation is suppressed,
that explains why the production cross section of  $e^+ e^- \to$ vector charmonia exhibits a dip around 4.26~GeV.

In this work, we extend our coupled-channel formalism to investigate other $J^{PC}=1^{--}$ exotic states.
The purpose is to build up a heavy quark spin multiplet of charmed meson molecules~\cite{Wang:2014wga}.
In what follows, we do not aim to fully explain the production or decay patterns of these exotic hadrons.
Instead, we will investigate the molecular components in experimentally observed $J^{PC}=1^{--}$ exotic states.
In unquenched quark model (UQM), the wave functions encapsulate both long- and short-distance information,
and we intend to utilize them to explore the structure of exotic $Y$ and $\psi$ resonances.

A brief introduction of our coupled-channel framework is given in Sec.~\ref{framework},
where some subtleties of the HQSS are also discussed.
Section~\ref{results} is devoted to the analysis of our results and their interpretations.
In Sec.~\ref{decays}, a few remarks on the strong decays of \xyy/ and \ppsi/ are provided,
followed by a summary of this study in the Sec.~\ref{summary}.

\section{Coupled-Channel Formalism} \label{framework}

The quenched quark model described the spectrum of the low-lying states reasonably well,
but its predictions for the states nearby or above open-flavor thresholds are questionable. 
It lacks the influence of ``sea-quarks".
Heavy quarkonium can couple to intermediate 
heavy mesons through the creation of light quark-antiquark pair.
This enlarges the Fock space of heavy quarkonium and manifests
the presence of multiquark components in its wave function.
Such components will change the Hamiltonian of the potential model,
causing a mass shift due to self-energy corrections, and may also 
give direct contributions to strong and electromagnetic decays~\cite{Lu:2016mbb,Anwar:2018yqm}.
The probability of such multiquark (molecular) components can be worked out
(see e.g., Ref.~\cite{Lu:2017yhl}).

In UQM, a physical or experimentally observed hadron $\ket{A} $ can be expressed as
\be
\ket {A}=c_0 \ket{\psi_0} +\sum_{BC} \int d^3p\, c_{BC}(p) \ket{BC;p},
\ee
where $c_0$ and $c_{BC}$ stand for the normalization constant of the bare state 
and the $BC$ components, respectively.
In this work, $B$ and $C$ refer to charmed and anti-charmed mesons,
and the summation over $BC$ is carried out up to ground state $P$-wave charmed mesons ~\cite{Lu:2017hma}.
The bare state $\ket {\psi_0}$ is normalized to 1, and the physical state $\ket {A}$ is also normalized to 1
if it lies below $D\bar{D}$ threshold. $\ket{BC;p}$ is normalized as $\inner{BC;p_1}{B'C';p_2}=\delta^3(p_1-p_2)\delta_{BB'}\delta_{CC'}$, where $p$ is the momentum of meson $B$ in $\ket{A}$'s rest frame.
The effects from the $BC$ components are referred to as \cce/.

The full Hamiltonian of the physical state reads as
\be
\label{hamiltonian}
  H=H_0+H_{BC}+H_I,
\ee
where $H_0$ is the Hamiltonian of the bare state (see Appendix~\ref{bareH0} for details), the continuum Hamiltonian is $H_{BC} \ket{BC;p} = E_{BC}\ket{BC;p}$ with $E_{BC}=\sqrt{m_B^2+p^2}+\sqrt{m_C^2+p^2}$ is the energy of the continuum state (the interaction between $B$ and $C$ mesons is neglected here and transition between one continuum to another is not included), and $H_I$ is the interaction Hamiltonian which triggers the mixing of the bare state to the continuum.

For the bare-continuum mixing, which is an important dynamical pieces of the UQM, we adopt the
widely used ${}^3P_0$ model~\cite{Micu:1968mk}. In this model, the nonperturbative creation of light 
quark-antiquark pairs is triggered from the vacuum of quantum chromodynamics (QCD) with $J^{PC}=0^{++}$,
which in the spectroscopical notation ${}^{2S+1}L_J$ can be written as $\3P0$~\cite{Lu:2016mbb,Ferretti:2020civ}.
The interaction Hamiltonian can be expressed as
\be
H_I=2 m_q \gamma \int d^3x \bar{\psi}_q \psi_q,
\ee
where $m_q$ is the produced quark mass,
and $\gamma$ is the dimensionless coupling constant.
The $\psi_q$ ($\bar{\psi}_q$) is the spinor field to generate anti-quark (quark).
Since the probability to generate heavier quarks is suppressed,
we use the effective strength $\gamma_s=\frac{m_q}{m_s}\gamma$ in the following calculation,
where $m_q=m_u=m_d$ is the constituent quark mass of up (or down) quark and $m_s$  is strange quark mass.
Their numerical values are listed in Appendix~\ref{bareH0} (Table~\ref{tab:4260Para}).

The mass shift caused by the $BC$ components and their probabilities are obtained after
solving the Schr\"{o}dinger equation with the full Hamiltonian $H$.
They are expressed as
\bal
\Delta M&:=M-M_0=\sum_{BC}\int d^3p\, \frac{\vert \bra {BC;p} H_I \ket{\psi_0} \vert ^2}{M-E_{BC}-i\epsilon}, \label{massShift}\\
P_{BC}&:=\int d^3p|c_{BC}|^2=\int d^3p\, \frac{\vert \bra {BC;p} H_I \ket{\psi_0} \vert ^2}{(M-E_{BC})^2},\label{PBC}
\eal
where $M$ and $M_0$ are the eigenvalues of the full ($H$) and quenched/bare Hamiltonian ($H_0$), respectively;
$P_{BC}$ is the unnormalized probability, which is also called the coupling strength in the next section.
In order to analyze different partial-wave contributions,
we adopt the Jacob-Wick formula to separate different partial waves of $P_{BC}$~\cite{Jacob:1959at}.
(See Ref.~\cite{Lu:2016mbb} for a derivation of the above relations and UQM calculation details.)

To proceed, we need to specify that the coupled-channel calculation cannot be pursued if the wave functions
of the $\ket{\psi_0}$ and $BC$ components are not settled in Eqs.\,(\ref{massShift})$-$(\ref{PBC}).
The major part of the coupled-channel calculation is encoded in the wave function overlap integration,
\bal
 \bra{BC;p} H_I \ket{\psi_0}&= \int d^3k~
\phi_0(\vec{k}+\vec{p}) ~\phi_B^*(\vec{k}+x \vec{p}) \\ \nonumber
  &\times \phi_C^*(\vec{k}+x \vec{p})~ |\vec{k}| ~Y_1^m(\theta_{\vec{k}},\phi_{\vec{k}})
  \eal
where $x=m_q/(m_Q+m_q)$, and $m_Q$ and $m_q$ denote the charm quark and the light quark mass, respectively.
The $\phi_0, \phi_B$ and $\phi_C$ are the wave functions of $\ket{\psi_0}$ and $BC$ components, respectively and the notation $*$ stands for the complex conjugate.
These wave functions are in momentum space, and they are obtained by the Fourier transformation of the eigenfunctions of the bare Hamiltonian $H_0$.

For the heavy quarkonium decays, the heavy quarks are treated as spectators in the \PPP/ model.
Their polarizations will not change after the generation of the light quark-antiquark pairs.
This leads us to conclude that the \PPP/ model itself respects the HQSS.
Nevertheless, some HQSS breaking effects can still slip into the coupled-channel calculation.
The breaking effects mainly lie in the input of the charmed mesons masses
and can be noticed from their mass splitting in the same $j_l^P$ multiplet,
where $j_l$ is the total spin of the light-quark cloud (with parity $P$) in the charmed meson.
For example, $D_2^*$ and $D_1$ belong to the same $j_l^+=3/2$ multiplet, but they do not degenerate
experimentally as claimed by the HQSS.

The HQSS further leads to configuration mixings of charmed mesons having different $j_l$.
For example, the experimentally observed $D_1(2420)$ (or $D_1$ in short) and $D_1(2430)$ (or $D_1'$)
are not the pure $j_l^+=3/2$ and $j_l^+=1/2$ (or the quark model ${}^3P_1$ and ${}^1P_1$ ) states, respectively,
but are their linear combinations.
Therefore, the wave functions from the quark model should be modified accordingly.
The mixture can be formulated as
\be\label{HQSSMixing}
\left(
  \begin{array}{c}
   \ket{D_1}\\
   \ket{D_1'} \\
  \end{array}
  \right)=
\left(
  \begin{array}{cc}
    \cos\, \theta & \sin\, \theta\\
    -\sin\, \theta& \cos\, \theta\\
  \end{array}
  \right)
  \left(
  \begin{array}{c}
   \ket{{}^3P_1}\\
   \ket{{}^1P_1}\\
  \end{array}
  \right),
\ee
where $\theta$ is the mixing angle.
The HQSS predicts it to be $\theta_0=\arctan(\sqrt{2})\approx 54.7^\circ$,
which is called the ideal mixing angle~\cite{Close:2005se,Qin:2016spb}.
Experimentally, the deviation from ideal mixing is observed ($5.7^\circ\pm 4^\circ$)~\cite{Belle:2003nsh},
which is very small compared with $\theta_0\approx 54.7^\circ$.
As reported earlier~\cite{Lu:2017yhl}, our results have minor changes if this deviation is added to $\theta_0$.
Hence, for simplicity, we will use ideal mixing of HQSS in this study.

Our bare mass value for the $\psi(3D)$ is nearly 250~MeV higher than the \xyy/, as listed in Table~\ref{tab:4260Para}.
Hadron loops can shift bare mass down to \xyy/ to compensate such a large mass gap~\cite{Barnes:2007xu}.
This requires the fitting of the charmonium spectrum which involves much {\it ab initio} calculations
(such as fixing of \PPP/'s coupling constant $\gamma$) which is beyond the scope of this work.
Therefore, $\gamma$ is not fixed here, that explains why $P_{BC}$ in Eq.~(\ref{PBC}) is unnormalized.
As a consequence, the absolute probabilities of the various \dmeson/ components are not determined.
Instead, by analyzing the $P_{BC}$, one can still draw very useful conclusions~\cite{Lu:2017yhl},
as demonstrated in the following sections.

\section{Results and Discussions} \label{results}

\subsection{Coupling Strengths for \xyy/}

We compute the probabilities of intermediate charmed meson components by
solving the Schr\"{o}dinger equation with the full Hamiltonian $H$ of Eq.\,(\ref{hamiltonian}).
Note that the \xyy/ lies in the mass range of $\psi(3S) \sim \psi(4S)$ and $\psi(2D) \sim \psi(3D)$
quark model states~\cite{Godfrey:1985xj,Segovia:2008zz}. To compare the results under different assumptions,
we calculate the probabilities for all four aforementioned bare states.
For consistency, the same set of parameters is being adopted which was previously
used in the study of \xy/~\cite{Lu:2017yhl}.

The unnormalized coupling strengths of \xyy/ with the charmed meson components are given in Table~\ref{unProb}.
Due to higher mass of \xyy/, we only consider those channels which involve at least
one $P$-wave charmed meson. It is interesting to notice that one of the channels with largest coupling to \xyy/ is \donedstar/.
The channel \dtwodstar/ also shows a sizable coupling with \xyy/ for $\psi(3S)$ and $\psi(4S)$ cores.
However, due to several reasons, discussed in the following, \xyy/ can be regarded as having
a dominant \donedstar/ component.

For the coupled channels which involve one $S$- and one $P$-wave charmed meson,
the HQSS leads to set $M_D=M_{D^*}$ ($j_l^-=1/2$ doublet),  $M_{D_0^*}=M_{D'_1}$ ($j_l^+=1/2$ doublet),
and $M_{D_1}=M_{D_2^*}$ ($j_l^+=3/2$ doublet).
As a result, the contribution of all those channels which involve charmed meson form the same $j_l$ multiplet
will be of the same order [except for the $D-D_{1}(D_{1}^{'})$ and \dtwod/ channels, which are the decay channels of \xyy/].
In such a scenario, the largest coupling will naturally come from the \dtwodstar/ channel,
since the spin configurations for this channel are at the maximum.
We refer this as \textit{spin enhancement mechanism}.
However, if the physical masses are applied to these \dmeson/s,
the \dtwodstar/ channel will be a little further from the \xyy/ and is expected to have
a smaller coupling strength.

For \xyy/, we noticed that the spin-enhancement mechanism is giving sizable contribution when we use the experimental mass of \xyy/ as the PDG average value $4368 \pm 13$ MeV. However, a recent analysis by the BESIII Collaboration for the process $e^+ e^- \to \pi^+ \pi^- \psi (2S)$ reported a larger mass of this resonance as $4383.8 \pm 4.2 \pm 0.8$~\cite{BESIII:2017tqk}, and an independent analysis of the BESIII combined data of $e^+ e^- \to \pi^+ \pi^- J/\psi$ and $e^+ e^- \to \pi^+ \pi^- \psi (2S)$ extracted the mass of \xyy/ resonance as $4386.4 \pm 2.1 \pm 6.4$~ \cite{Zhang:2017eta}. This indicates that \xyy/ might have a larger mass. The coupling of \xyy/ with different charmed meson channels found to be sensitive to its mass. With the use of a larger mass, the coupling of \xyy/ to the channel \donedstar/ dominantly exceeds the \dtwodstar/ channel and makes it a more reasonable candidate for a \donedstar/ molecule.

The couplings to those channels that are far above 4.368 GeV are generally small\textemdash a universal conclusion from the \cce/.
The asymptotic behavior of $P_{BC}$ is proportional to $1/(m_B+m_C)^2$.
If the coupled channels are further from \xyy/, their contributions will be naturally suppressed; we call this the \textit{mass-suppression mechanism}. The emergence of this mechanism can be seen from Table~\ref{unProb}, where the higher channels have very small contributions. 

For a solid conclusion,
the two mechanisms\textemdash mass suppression and the spin enhancement\textemdash 
have to compete with each other to tell us which channel gives the dominant contribution.
We will come to this point in the next subsection.
If the $D_s$ mesons are involved in the coupling channels, then an additional suppression comes from the effective strength of the \PPP/ model $\gamma_s$. Since $\gamma_s\approx0.66 \gamma$ (using the constituent quark mass from Appendix~\ref{bareH0}), the couplings to $D_s$ mesons are universally smaller than the $D$ mesons.

It is worth mentioning that, even though the non-\donedstar/ components are suppressed
when the charmonium core is $\psi(nD)$, the contribution of these components could still be sizable.
As shown in Table~\ref{unProb},
for the $\psi(3D)$, the contribution from the \dtwodstar/ channel is still around the \donedstar/ one.
However, the mass prediction for the quark model $\psi(3D)$ state is nearly $200$ MeV $-$ $300$ MeV above
the \xyy/~\cite{Godfrey:1985xj}; thus, having the $\psi(3D)$ core in its wave function is not well justified.
The most likely possibility for the charmonium core of the \xyy/ is $\psi(2D)$ due to the following reasons:

\begin{enumerate}
 
 \item The production cross section of vector charmonia in $e^+e^- \to$ hadrons (the $R$-value)
 is proportional the wave function at the origin, which is zero for the case of $\psi(nD)$.
Since the $\psi(nD)$ can only couple to the virtual photon at the next-to-next-to leading order \cite{Rosner:2001nm},
its direct production at the $e^+e^-$ collider is suppressed.
 
\item The $R$-value measured by the BES Collaboration~\cite{BES:2007zwq} 
has a dip instead of a peak around $4.368$ GeV, which indicates that the \xyy/ is likely to have a $D$-wave
charmonium core, and its mass value makes $\psi(2D)$ the most promising.

 \item The coupling of $\psi(2D)$ with the \donedstar/ is found to be the maximum, in two different sets of parameters and wave functions (see details in Table~\ref{nProb}). In this sense, our results support the dominant long-distance component of \xyy/ to be the \donedstar/.
 
\end{enumerate}

This enables us to conclude that the structure of \xyy/ is a mixture of a charmonium core $\psi(2D)$ with the dominant molecular component \donedstar/.

\begin{table}
  \renewcommand\arraystretch{1.4}
  \setlength{\tabcolsep}{11pt}
  \centering
   \begin{tabular}{ccccc}
  \hline\hline
  Channels & $\psi(3S)$ & $\psi(4S)$ & $\psi(2D)$ & $\psi(3D)$ \\
  \hline
  $D_1(D_1')-\bar{D}$          & 0       & 0     & 0     & 0     \\
  $D_{1}^{'}-\bar{D}^{*}$     & 1.580   & 1.254 & 0.910 & 0.707 \\
  $D_{1}-\bar{D}^{*}$             & 4.401   & 2.189 & 4.319 & 1.766 \\
  $D_{1}^{'}-\bar{D}_{1}^{'}$  & 0.604   & 0.227 & 0.738 & 0.300 \\
  $D_{0}^{*}-\bar{D}_{1}^{'}$  & 0.302   & 0.085 & 0.078 & 0.023 \\
  $D_{1}-\bar{D}_{1}^{'}$     & 0.511   & 0.214 & 0.274 & 0.151 \\
  $D_{2}^{*}-\bar{D}_{1}^{'}$    & 0.785   & 0.519 & 0.514 & 0.353 \\
  $D_{1}-\bar{D}_{0}^{*}$   & 0.219   & 0.081 & 0.122 & 0.049 \\
  $D_{1}-\bar{D}_{1}$         & 0.631   & 0.298 & 0.632 & 0.314 \\
  $D_{2}^{*}-\bar{D}_{1}$      & 1.010   & 0.382 & 0.788 & 0.326 \\
  $D_{2}^{*}-\bar{D}$          & 0   & 0 & 0 & 0 \\
  $D_{2}^{*}-\bar{D}^*$       & 6.110   &  2.938 & 3.468 & 1.935 \\
  $D_{2}^{*}-\bar{D}_{0}^{*}$        & 0.655    & 0.213 & 0.505 & 0.212 \\
  $D_{2}^{*}-\bar{D}_{2}^{*}$        &  1.151  & 0.915 & 1.214 & 0.988 \\
  \hline\hline
\end{tabular}
\caption{The unnormalized coupling strength of various channels with the initial state mass $=4368$ MeV for the different charmonium states in the HQSS limit.}
\label{unProb}
\end{table}

To cross check the model dependence of our results and the validity of our conclusions, 
we try different sets of parameters and different wave-function approximations.
For this purpose, we adapt the quark model parameters from Barnes and Swanson~\cite{Barnes:2007xu} and
simple harmonic oscillator (SHO) approximations for the wave functions.
The results for the normalized coupling strengths are compared in Table~\ref{nProb}.
The wave functions used by Barnes and Swanson~\cite{Barnes:2007xu}
are SHO approximations~\cite{Anwar:2016mxo} but are useful to cross check the coupling pattern.
One of the largest coupling channels to \xyy/ is \donedstar/ even if we use different set of parameters.
This conclusion, to some extent, is model independent, since we tried two different sets of parameters
and different wave functions to cross check the model dependence.

\begin{table*}
  \renewcommand\arraystretch{1.4}
  \setlength{\tabcolsep}{12pt}
  \centering
\begin{tabular}{l|llll|lllllll}
\hline\hline
\multirow{2}*{Coupled Channels}&\multicolumn{4}{c|}{Benchmark-I} &\multicolumn{4}{c}{Benchmark-II} \\
 \cline{2-9}
& $\psi(3S)$ & $\psi(4S)$ & $\psi(2D)$ & $\psi(3D)$ & $\psi(3S)$ & $\psi(4S)$ & $\psi(2D)$ & $\psi(3D)$\\
\hline
$D_1(D_1')-\bar{D}$ & 0     & 0     & 0     & 0          & 0     & 0     & 0     & 0     \\
 $D_{1}^{'}-\bar{D}^{*}$   & 0.269 & 0.412 & 0.132 & 0.256      & 1.197  & 1.529 & 0.426 & 0.874 \\
 $\textit{\textbf{D}}_{\textbf{1}}-  \bar{\textit{\textbf{D}}}^{\textbf{*}}$&
 \textbf{1} & \textbf{1} & \textbf{1} & \textbf{1} & \textbf{1} & \textbf{1} & \textbf{1} & \textbf{1} \\
 $D_{1}^{'}-\bar{D}_{1}^{'}$ & 0.103 & 0.075 & 0.107 & 0.109      & 0.193  & 0.354 & 0.254 & 0.456  \\
 $D_{0}^{*}-\bar{D}_{1}^{'}$ & 0.051 & 0.028 & 0.011 & 0.008      & 0.069 & 0.122 & 0.027 & 0.051 \\
 $D_{1}-\bar{D}_{1}^{'}$     & 0.087 & 0.07  & 0.04  & 0.054      & 0.165 & 0.305 & 0.086  & 0.216 \\
 $D_{2}^{*}-\bar{D}_{1}^{'}$ & 0.133 & 0.170 & 0.075 & 0.128      & 0.416 & 0.762 & 0.234 & 0.564  \\
 $D_{1}-\bar{D}_{0}^{*}$   & 0.037 & 0.027 & 0.018 & 0.018      & 0.059 & 0.111 & 0.042 & 0.087 \\
 $D_{1}-\bar{D}_{1}$    & 0.107 & 0.098 & 0.092 & 0.114      & 0.240  & 0.432 & 0.255 & 0.448 \\
 $D_{2}^{*}-\bar{D}_{1}$   & 0.171 & 0.125 & 0.115 & 0.118      & 0.333 & 0.629 & 0.237 & 0.515 \\
 $D_{2}^{*}-\bar{D}$    & 0 & 0 & 0 & 0      & 0 & 0 & 0 & 0  \\
 $D_{2}^{*}-\bar{D}^*$            & 1.386 & 1.342 & 0.803 & 1.096      & 1.520 & 1.571 & 0.809 & 1.270  \\
 $D_{2}^{*}-\bar{D}_{0}^{*}$  & 0.144 & 0.096 & 0.114 & 0.118       &  0.188 & 0.355 & 0.147 & 0.368  \\
 $D_{2}^{*}-\bar{D}_{2}^{*}$   & 0.171 & 0.209 & 0.141 & 0.280       & 0.379 & 0.665 & 0.281 & 0.658  \\
\hline\hline
\end{tabular}
\caption{Coupling strength of various coupled channels normalize to \donedstar/ with the initial state mass $=4368$ MeV.
  Benchmark-1 corresponds to our parametrization while Benchmark-II refers to
  parametrization used by Barnes and Swanson~\cite{Barnes:2007xu}.
  One should not compare the numbers between different columns until \donedstar/ values in Table~\ref{unProb} are multiplied.}
\label{nProb}
\end{table*}

\begin{figure}
  \centering
  \includegraphics[width=0.5\textwidth]{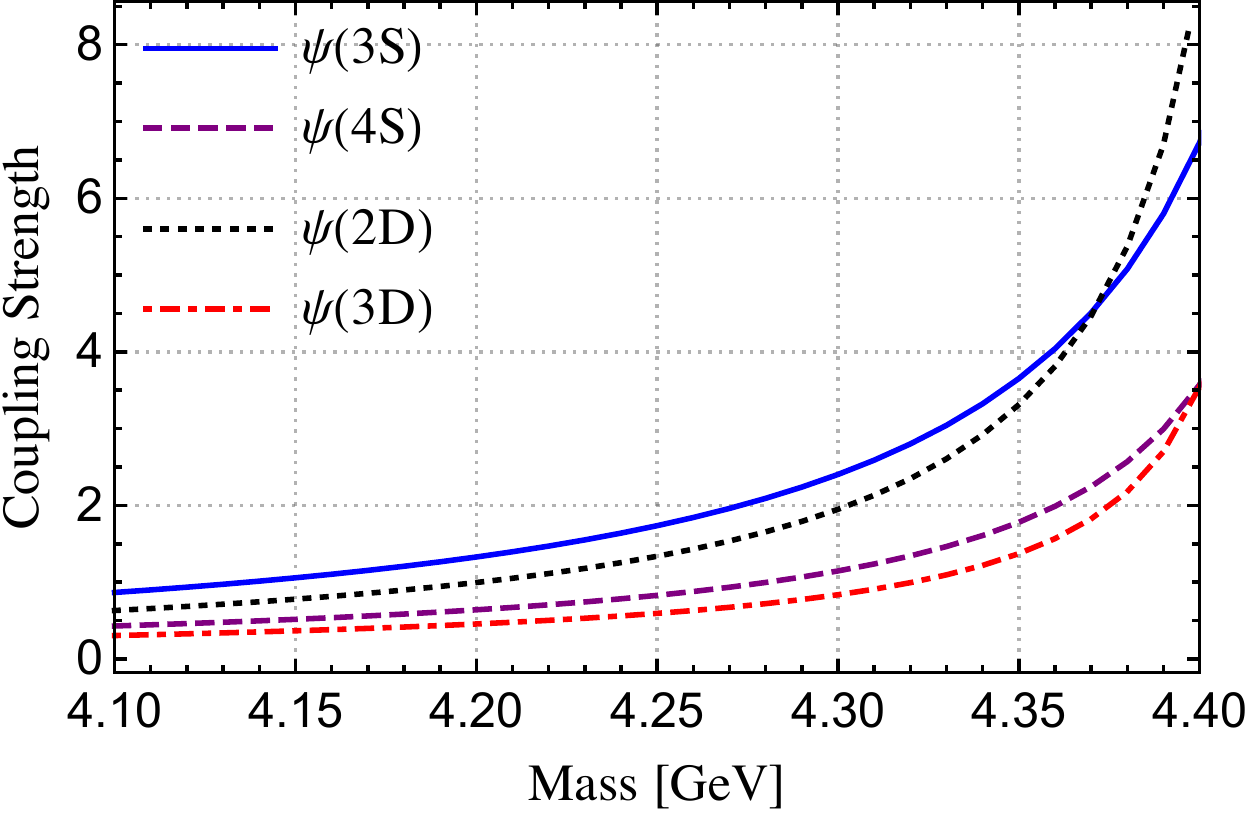}\\
  \caption{Coupling strength of the \donedstar/ channel with the \xyy/ as a function of its mass for the different charmonium wave functions.}
\label{D1DMassplot.pdf}
\end{figure}

The behavior of the \donedstar/ coupling strength as a function of mass of
the \xyy/ resonance is shown in Fig.~\ref{D1DMassplot.pdf}.
We try different initial charmonium wave functions and push the initial mass
of the resonance to vary in a vast energy range up to 4.4 GeV.
A stable behavior of the coupling is found.
The coupling gets enhanced as the initial mass of the resonance approaches
the \donedstar/ threshold, and this behavior is the same for all considered
initial charmonium wave functions. This verifies our conjecture, i.e.,
\xyy/ has the largest coupling to the \donedstar/ channel in all wave-function
choices. Hence, we can conclude that \xyy/ can be described in a molecular
picture having \donedstar/ as its largest component, and thus can be interpreted as a
spin partner of the \xy/.

\subsection{ $\psi(4415)$ and \dtwodstar/ Channel}

\begin{figure}
  \centering
  \includegraphics[width=0.5\textwidth]{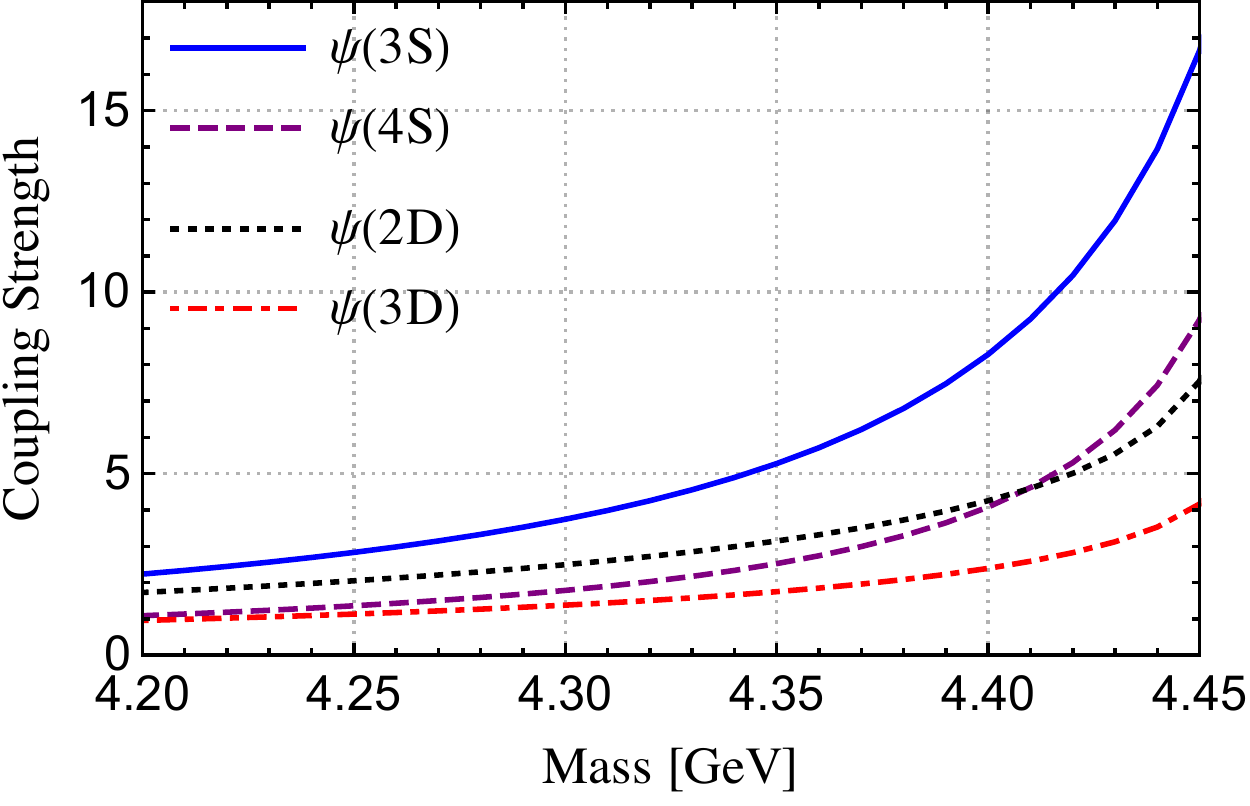}\\
  \caption{Coupling strength of \dtwodstar/ channel with the \ppsi/ as a function of its mass for the different charmonium wave functions.}
\label{D2DMassplot.pdf}
\end{figure}

For the case of $\psi(4415)$,
since it is extremely close to the \donedstar/ threshold, one can expect that
this state will show dominant coupling to the aforementioned channel.
However, it is found that the spin-enhancement mechanism takes over here and the $\psi(4415)$ has shown the largest coupling
to the \dtwodstar/ channel (see Table~\ref{psi4415}).
The coupling of this channel with the $\psi(4S)$ and $\psi(2D)$ cores is almost the same.
The behavior of the \dtwodstar/ coupling strength as a function of mass of
the \ppsi/ resonance is shown in Fig.~\ref{D2DMassplot.pdf}.

Since the next higher channels above \dtwodstar/, such as $D_2^*\bar{D}^*_0$ or $D^*_2 \bar{D}_2^*$
(both of these channels couple to $J^{PC}=1^{--}$ in $P$-wave),
are further from the \ppsi/, and the mass suppression mechanism implies that the
contributions from all these higher components will be highly suppressed,
which makes the name of the \dtwodstar/ molecule more reasonable.

HQSS also implies the presence of other molecular component in the wave function of \ppsi/
with different heavy quark spin structure such as \doned/ and \dtwod/;
the latter channel couples to $J^{PC}=1^{--}$ through $D$-wave only.
However, these channels are open for \ppsi/ because the mass of these 
thresholds are below the experimental mass of \ppsi/.
This is merely the reason for zero coupling strengths in the Table~\ref{psi4415}. 

An important observation is that,
if the charmonium core of the \ppsi/ is in the $S$-wave,
the two mechanisms (mass suppression and the spin enhancement) will be of similar importance.
As a consequence, non-negligible contributions other than the \dtwodstar/ channel may exist.
This indeed showed up in our results and can be seen from Table~\ref{psi4415}.
The coupling strength of \ppsi/ to the \donedstar/  is very close to the \dtwodstar/  channel
for the case of $\psi(3S)$ and $\psi(4S)$. 
In such a scenario, one can argue that the $\psi(4415)$ is dominant by \donedstar/ rather than \dtwodstar/.
However, the sizable coupling of the $\psi(4415)$ with the channel \donedstar/ is merely
an artifact of just being extremely close to this threshold.

It is most likely that the charmonium core of $\psi(4415)$ is in $S$-wave which results the same order of coupling for all initial core wave functions, as can be seen from Table~\ref{psi4415}. This is further supported by the $R$-value measurement that the production cross section is enhanced significantly around $\psi(4415)$'s mass which normally is the case for $\psi(nS)$ cores. The mass difference between $\psi(4415)$ and the \dtwodstar/ threshold is just close enough to have the same binding energy as \xy/ which reflects an important consequence of HQSS.

This analysis leads us to conclude that $\psi(4415)$ is likely to be a mixture of a short-distance core $\psi(nS)$ and the dominant long-distance \dtwodstar/ component. However, coupling strengths alone are not enough to conclude the structure of these resonances. An important way to disentangle the molecular configuration of \xyy/ and $\psi(4415)$ is to look for suggested decay patterns, as discussed in the following section.

\begin{table}
  \renewcommand\arraystretch{1.4}
  \setlength{\tabcolsep}{11pt}
  \centering
   \begin{tabular}{ccccc}
  \hline\hline
  Channels & $\psi(3S)$ & $\psi(4S)$ & $\psi(2D)$ & $\psi(3D)$ \\
  \hline
  $D_1(D_1')-\bar{D}$          & 0       & 0     & 0     & 0     \\
  $D_{1}^{'}-\bar{D}^{*}$      & 2.129   & 3.768 & 3.811 & 6.335 \\
  $D_{1}-\bar{D}^{*}$            & 10.324   & 5.937 & 1.157 & 1.256 \\
  $D_{1}^{'}-\bar{D}_{1}^{'}$  & 0.653   & 0.240 & 0.176 & 0.085 \\
  $D_{0}^{*}-\bar{D}_{1}^{'}$  & 0.333   & 0.090 & 0.097 & 0.041 \\
  $D_{1}-\bar{D}_{1}^{'}$      & 0.556   & 0.228 & 0.130 & 0.048 \\
  $D_{2}^{*}-\bar{D}_{1}^{'}$  & 0.844   & 0.556 & 0.151 & 0.107 \\
  $D_{1}-\bar{D}_{0}^{*}$      & 0.243   & 0.087 & 0.058 & 0.017 \\
  $D_{1}-\bar{D}_{1}$          & 0.685   & 0.319 & 0.162 & 0.086 \\
  $D_{2}^{*}-\bar{D}_{1}$      & 1.087   & 0.401 & 0.295 & 0.144 \\
  $D_{2}^{*}-\bar{D}$         & 0   & 0 & 0 & 0 \\
  $D_{2}^{*}-\bar{D}^*$      & 10.594   &  5.381 & 5.057 & 2.851 \\
  $D_{2}^{*}-\bar{D}_{0}^{*}$      & 0.762    & 0.233 & 0.561 & 0.227 \\
  $D_{2}^{*}-\bar{D}_{2}^{*}$      &  1.307  & 1.039  & 1.348 & 1.100 \\
  \hline\hline
\end{tabular}
\caption{The unnormalized coupling strength of various channels around $\psi(4415)$ mass for the different charmonium states in the HQSS limit.}
\label{psi4415}
\end{table}

\section{Remarks on Decays} \label{decays}

The open-charm decays of \xyy/ provide an important pathway to disentangle its long-distance structure.
The natural decay of a molecular state is into its constituents~\cite{Guo:2017jvc}, and the strongest decay channel of the $D_1$ is $D^*\pi$.
Hence, in the \donedstar/ molecular configuration, we argue that the dominant decay mode of the \xyy/ is likely to be the $D^* \bar{D}^* \pi$,
and the decay of \xyy/ into the $D \bar{D}^* \pi$ final state is not possible. This is the reason that the \xyy/ has not shown up in the
$e^+ e^- \to D \bar{D}^* \pi$ final state at Belle~\cite{Belle:2009dus}. 

In fact, the $D \bar{D}^* \pi$ final state 
is expected to be the dominant decay mode of the \xy/ in the molecular picture~\cite{Cleven:2013mka},
which is supported by the recent precise measurements of BESIII Collaboration~\cite{BESIII:2018iea}.
It is interesting to mention that the \xyy/ is again invisible in this recent BESIII data~\cite{BESIII:2018iea}.
Hence, it is very important to measure $e^+ e^- \to D^* \bar{D}^* \pi$ cross sections
to prove/exclude the existence (as a hadronic molecule) of \xyy/ which we claim is a heavy quark spin partner of the \xy/.

In contrast to above, the $D_2^*$ meson can decay into $D^*\pi$ and $D\pi$. Therefore, a \dtwodstar/ molecular state must leave strong imprints in the $D \bar{D}^* \pi$ and $D^* \bar{D}^* \pi$ final states.  For $\psi(4415)\to  D \bar{D}^* \pi$, an upper limit for the partial decay width has been extracted by the Belle to be smaller than $11\%$\cite{ParticleDataGroup:2020ssz,Belle:2009dus}.
The broad enhancement around $4.40$ GeV in the recent BESIII measurement of $e^+ e^- \to D \bar{D}^* \pi$~\cite{BESIII:2018iea} indicates clear contributions from \ppsi/. If the same pattern is observed in the $e^+ e^- \to D^* \bar{D}^* \pi$ process, it will indicate that the \ppsi/ is likely to have a dominant \dtwodstar/ component which would provide a quantitative support to our conclusion.

The precise evaluation of the decay widths of the above discussed decays requires an accurate knowledge of the coupling of a molecular state with its components. One can benefit from the recent BESIII data~\cite{BESIII:2018iea} to extract the \xy/$-$\doned/ effective coupling and use HQSS to relate it to other molecular configurations. This would be the topic of our future explorations~\cite{anwar&dong}.

\section{Summary} \label{summary}

We computed and analyzed the probabilities of various \dmeson/ molecular components for \xyy/
under the coupled-channel formalism by assigning $\psi(3S)$, $\psi(4S)$, $\psi(2D)$ and $\psi(3D)$ initial wave functions.
We found that the channel \donedstar/ couples more strongly around the \xyy/ mass regime,
and the coupling behavior is the same for all considered initial charmonium wave functions.
This enlightens that the most favorable molecular scenario for \xyy/ is the \donedstar/,
and hence, it can be interpreted as a heavy quark spin partner of the \xy/.

We also analyze the coupling of \ppsi/ with several nearest \dmeson/ channels. 
It shows the largest coupling to the \dtwodstar/ channel in all four initial charmonium wave functions,
which makes \ppsi/ a good candidate for a prominent \dtwodstar/ molecule.
By this means, we argue that the heavy quark spin multiplet involving one $P$- and one $S$-wave
charmed meson is considered as completed.

However, the hitherto unobserved decays of these resonances are highly decisive and demanding,
such as $ \Psi \to D^* \bar{D}^* \pi$, where $\Psi \in \{Y(4360), \psi(4415)\}$.
Once the predicted pattern of heavy quark spin multiplet is confirmed by future experiments,
it will serve to deepen our understanding of how QCD forms hadronic matter by arranging multiquarks.

\begin{acknowledgments}

We are grateful to Christoph Hanhart for a careful reading of the manuscript and very enlightening remarks,
and to Feng-Kun Guo, Qian Wang, and Bing-Song Zou for several helpful discussions and suggestions.
This work is supported by the Deutsche Forschungsgemeinschaft (DFG) and the National Natural Science
Foundation of China (NSFC) through the funds provided to the Sino-German Collaborative Research Center TRR110
``Symmetries and the Emergence of Structure in QCD" (DFG Project-ID 196253076, NSFC Grant No. 12070131001).

\end{acknowledgments}

\appendix

\section{Bare Hamiltonian}
\label{bareH0}

Bare charmonium states are obtained by solving the Schr\"odinger
equation with the well-known Cornell potential~\cite{Eichten:1978tg,Eichten:1979ms},
which incorporates a spin-independent color Coulomb plus
linear confined (scalar) potential.
In the quenched limit, the potential can be written as
\be \label{pot1}
V(r)=-\frac{4}{3} \frac{\alpha}{r}+\lambda r+c,
\ee
where $\alpha, \lambda$ and $c$ stand for the strength of the color Coulomb potential,
the strength of linear confinement, and mass renormalization, respectively.
The hyperfine and fine structures are generated by the spin-dependent interactions,
 \bal \label{pot2}
  V_{s}(r)&= \frac{1}{m^2_c}\bigg[ \left(\frac{2\alpha_s}{r^3}-\frac{\lambda}{2 r}\right)\vect{L}\cdot \vect{S}
  +\frac{32\pi \alpha_s}{9}~\tilde{\delta}(r)~\vect{S}_c\cdot \vect{S}_{\bar{c}}\\ \nonumber
  &+\frac{4\alpha_s}{r^3}\left(\frac{\vect{S}_c\cdot \vect{S}_{\bar{c}}}{3}
  +\frac{(\vect{S}_c\cdot\vect{r}) (\vect{S}_{\bar{c}}\cdot \vect{r})}{r^2}\right) \bigg]~,
  \eal
where $\vect{L}$ denotes the relative orbital angular momentum,
$\vect{S}=\vect{S}_c+\vect{S}_{\bar{c}}$  is the total spin of the charm quark pairs,
and $m_c$ is the charm quark mass.
The smeared $\tilde{\delta}(r)$ function can be read from Refs.~\cite{Barnes:2005pb,Li:2009ad}.
These spin dependent terms are treated as perturbations.

The Hamiltonian of the Schr\"{o}dinger equation in the quenched limit is represented as
\be\label{quenched}
H_0 =2m_c+\frac{p^2}{m_c}+V(r)+V_{s}(r).
\ee
The spatial wave functions and bare mass $M_0$ are obtained by
solving the Schr\"{o}dinger equation numerically using the Numerov method~\cite{Numerov:1927}.
We borrow the parameters of the potential model [Eqs.~(\ref{pot1},\ref{pot2})] from Ref.~\cite{Li:2009ad}
and the bare-mass spectrum predicted by $H_0$ is listed in Table~\ref{tab:4260Para} along with the model parameters.

\begin{table}
 \renewcommand\arraystretch{1.4}
  \setlength{\tabcolsep}{11pt}
  \centering
\begin{tabular}{cccc}
\hline\hline
$\alpha$ & $\lambda$ & $c$ & $\sigma$\\
0.55 & $0.175~\text{GeV}^2$ &$-0.419~\text{GeV}$ & 1.45~GeV \\
\hline
$m_c$ & $m_s$ & $m_u$ & $m_d$\\
1.7 & 0.5 & 0.33 & 0.33\\
\hline\hline
$\psi(1S)$ & $\psi(2S)$ & $\psi(3S)$ & $\psi(4S)$\\
3.112 & 3.755 & 4.194 & 4.562 \\
\hline
$\psi(1D)$ & $\psi(2D)$ & $\psi(3D)$ & $\psi(4D)$\\
3.878 & 4.270 & 4.613 & 4.926\\
\hline\hline
\end{tabular}
\caption{Parameters of Cornell potential model and the corresponding bare mass spectrum.
The units of mass are GeV.}
\label{tab:4260Para}
\end{table}

\section{Charmed Mesons Thresholds}
\label{masses}

Some relevant thresholds involving at least one $P$-wave charmed meson in the proximity of \xyy/ and \ppsi/ are listed in Table~\ref{thresholds}.

\begin{table}[!h]
  \renewcommand\arraystretch{1.5}
  \setlength{\tabcolsep}{20pt}
  \centering
   \begin{tabular}{cc}
  \hline\hline
   Mesons                 & Threshold (MeV) \\
  \hline
  $D_{1}\bar D ~( D_{1}^{'}\bar D)$                   & 4292 (4282)  \\
  $D^{*}_{2} \bar D $                   & 4331 \\
  $D_{1} \bar{D}^{*} ~ (D_{1}^{'} \bar{D}^{*})$           & 4432 (4422)    \\
  $D^{*}_{2} \bar{D}^{*}$                              & 4471            \\
  $D_{1} \bar{D}_{0}^{*} ~ (D_{1}^{'} \bar{D}_{0}^{*} )$  & 4765 (4755)   \\
  $D^{*}_{2} \bar{D}_0^*$                   & 4804  \\
  $D_{1} \bar{D}_{1}~(D_{1}^{'}\bar{D}_{1}^{'})$       & 4844 (4824)     \\
  $D_{2}^{*} \bar{D}_{1}~ (D_{2}^{*} \bar{D}_{1}^{'})$   & 4883 (4873)        \\
  $D^{*}_{2} \bar{D}^{*}_2$                              & 4922            \\
  \hline\hline
  \end{tabular}
  \caption{Important thresholds of charmed mesons in the proximity of \xyy/ and \ppsi/ using PDG values~\cite{ParticleDataGroup:2020ssz}.}
  \label{thresholds}
\end{table}

\end{document}